\documentclass[journal=jacsat,manuscript=article]{achemso}
\usepackage{chemformula} 
\usepackage[T1]{fontenc} 
\usepackage{graphicx}
\usepackage{dcolumn}
\usepackage{bm}
\usepackage{xfrac}%
\usepackage{color}
\usepackage{tabularx}%
\usepackage{hyperref}
\usepackage{xcolor,colortbl}
\usepackage{multirow}

\newcommand{\RCL}{$\alpha$-RuCl$_{3}$}

\newcommand{\CIO}{Cu$_2$IrO$_3$}
\newcommand{\NIO}{Na$_2$IrO$_3$}
\newcommand{\LIO}{$\alpha$-Li$_2$IrO$_3$}
\newcommand{\LRO}{Li$_2$RhO$_3$}

\newcommand{\ALIO}{Ag$_{3}$LiIr$_2$O$_6$}

\newcommand{\HLIO}{H$_{3}$LiIr$_2$O$_6$}
\newcommand{\CLIO}{Cu$_{3}$LiIr$_2$O$_6$}

\newcommand{\tg}{$t_{\mathrm {2g}}$}
\newcommand{\eg}{$e_{\mathrm {g}}$}

\author{Faranak~Bahrami}
\affiliation{Department of Physics, Boston College, Chestnut Hill, MA 02467, USA}
\author{Mykola~Abramchuk}
\affiliation{Department of Physics, Boston College, Chestnut Hill, MA 02467, USA}
\author{Oleg~I.~Lebedev}
\affiliation{Laboratoire CRISMAT, ENSICAEN-CNRS UMR6508, 14050 Caen, France}
\author{Fazel~Tafti}
\affiliation{Department of Physics, Boston College, Chestnut Hill, MA 02467, USA}
\email{fazel.tafti@bc.edu}
\title{Metastable Kitaev Magnets}

\begin{document}


\begin{abstract}
Nearly two decades ago, Alexei Kitaev proposed a model for spin-$1/2$ particles with bond-directional interactions on a two-dimensional honeycomb lattice which had the potential to host a quantum spin-liquid ground state. 
This work initiated numerous investigations to design and synthesize materials that would physically realize the Kitaev Hamiltonian. 
The first-generation of such materials, such as \NIO, \LIO, and \RCL, revealed the presence of non-Kitaev interactions such as the Heisenberg and off-diagonal exchange. 
Both physical pressure and chemical doping were used to tune the relative strength of the Kitaev and competing interactions; however, little progress was made towards achieving a purely Kitaev system.
Here, we review the recent breakthrough in modifying Kitaev magnets via topochemical methods that has led to the second-generation of Kitaev materials.
We show how structural modifications due to the topotactic exchange reactions can alter the magnetic interactions in favor of a quantum spin-liquid phase.
\end{abstract}

\pagebreak
\section{\label{sec:intro}Introduction}

Recently, the 4d/5d honeycomb layered materials have been vigorously studied due to their potential in realizing a quantum spin-liquid (QSL) ground state~\cite{PRL_Jackeli_Mott, PRL_Singh_A2IrO3, Nature_takagi_concept_2019, PRB_Plumb_RCL, PRB_Williams_Incommensurate, nano_wang_range_2020, Todorova_LRO_2011, PRB_Choi_Spin}. 
First introduced by Alexei Kitaev in 2006, the Kitaev model is an exactly solvable theoretical model with bond-dependent Ising interactions among spin-$1/2$ degrees of freedom on a two-dimensional (2D) honeycomb lattice, which is described by the Kitaev Hamiltonian: $\mathcal{H}=-\sum K_{\gamma}\mathbf{S_i}^{\gamma}\mathbf{S_j}^{\gamma}$~\cite{Kitaev_anyons_2006}.
The ground state of this system is magnetically frustrated and is predicted to be a QSL~\cite{Kitaev_anyons_2006}. 
The applications of a Kitaev QSL in quantum information and the possibility of realizing Majorana fermions have inspired numerous investigations on quasi-2D honeycomb materials~\cite{Nature_takagi_concept_2019,knolle_field_2019,chaloupka_kitaev-heisenberg_2010,PRL_Jackeli_Mott,kim_novel_2008}.

\begin{figure}[ht]
\includegraphics[width=0.45\textwidth]{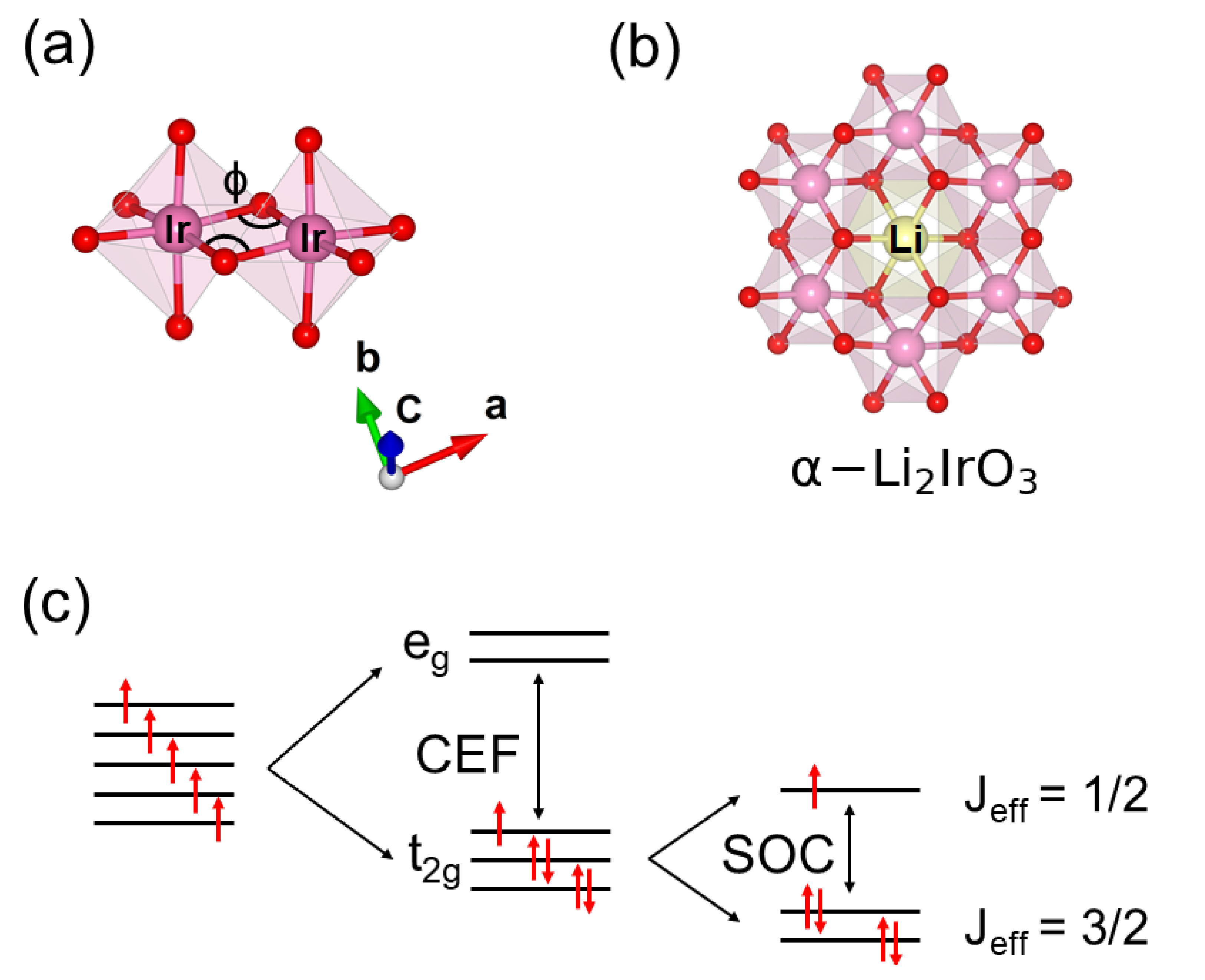}
\caption{\label{fig:T2G}
(a) The bond angle ($\phi$) between edge-shared octahedral units plays a significant role in tuning the magnetic interactions.
(b) Edge-sharing octahedral units create a honeycomb structure in Kitaev magnets such as \LIO\ and \NIO.
(c) Interplay between CEF and SOC creates the isospin-1/2 state in the Kitaev magnets.  
}
\end{figure}

The first-generation of these compounds, namely \NIO, \LIO, \LRO, and \RCL, were synthesized using conventional solid state methods at high temperatures ($T>700$~$^{\circ}$C).
In these materials, heavy transition metal ions (Ru$^{3+}$, Rh$^{4+}$, and Ir$^{4+}$) are octahedrally coordinated with oxygen or chlorine atoms (Fig.~\ref{fig:T2G}a), and the edge-sharing octahedra create honeycomb layers (Fig.~\ref{fig:T2G}b). 
The combination of octahedral crystal electric field (CEF) and strong spin-orbit coupling (SOC) splits the five-fold degenerate $d$-levels and leaves one electron in the isospin-$1/2$ ($J_{\text{eff}}$=$1/2$) state necessary for the Kitaev model (Fig.~\ref{fig:T2G}c)~\cite{PRL_Jackeli_Mott,PRB_Plumb_RCL, Todorova_LRO_2011, PRB_mehlawat_heat_2017,Kitaev_anyons_2006}. 

Finding new Kitaev magnets, beyond the first-generation compounds, has become a frontier challenge in solid state chemistry.
Prior attempts to replace Na with K in \NIO\ or replacing Cl with Br in \RCL\ have led to other stable phases with different structures instead of the honeycomb lattice~\cite{weber_trivalent_2017,merlino_orderdisorder_2004}.
The amount of physical pressure required to substantially tune the interactions is too high~\cite{bhattacharjee_spinorbital_2012} and chemical doping leads to a change of spin state~\cite{cao_challenge_2018}.
Therefore, recent success in synthesizing a second-generation of Kitaev magnets where magnetic interactions can be tuned by topochemical methods has revitalized the field.
In this review, we will first explain the different types of exchange reactions (partial and complete), then discuss the interplay between topochemical reactions and magnetism, and finally present heat capacity and magnetization data to compare the properties of the first and second-generation Kiteav magnets.

\section{\label{sec:topotactic}Topotactic Exchange Reactions}
\begin{figure}[ht]
\includegraphics[width=0.45\textwidth]{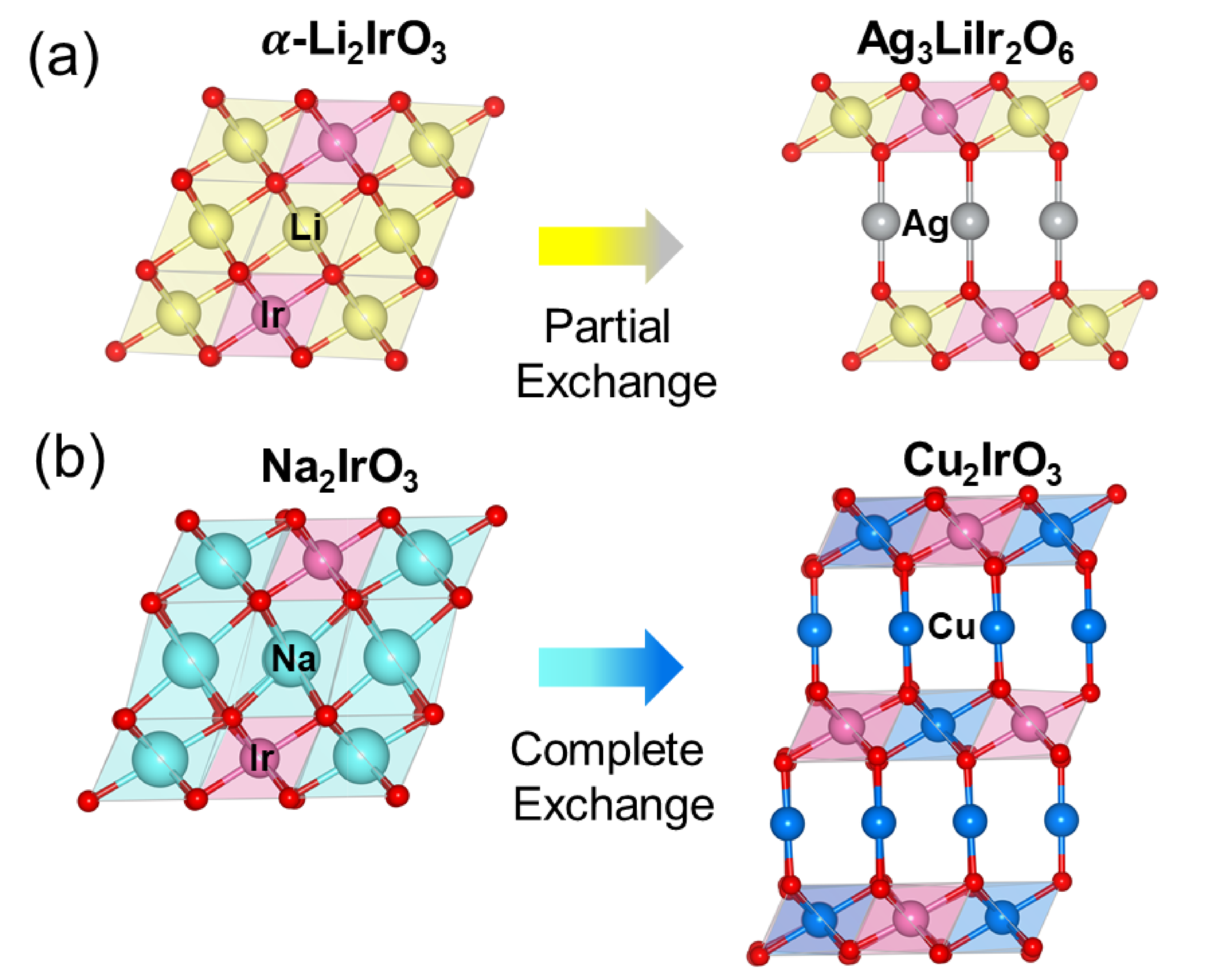}
\caption{\label{fig:Synthesis}
Synthesis of the second-generation Kitaev magnets from the first-generation materials through (a) partial and (b) complete exchange reactions.
Both generations have honeycomb layers.
The topochemical change of inter-layer coordination from octahedral to linear modifies the intra-layer Ir-O-Ir bond angles due to the change of oxygen positions.  
}
\end{figure}
The second-generation Kitaev magnets are metastable compounds, i.e. they have a higher enthalpy of formation and a lower decomposition threshold compared to stable counterparts~\cite{aykol_thermodynamic_2018}. 
Thus, it is impossible to synthesize them with conventional solid state methods at high temperatures.
Instead, they are stabilized through topochemical reactions from the first-generation compounds under mild conditions.
As shown schematically in Figs.~\ref{fig:Synthesis} the global symmetries of the unit cell (space group and honeycomb structure) do not change during a topochemical reaction.
However, the local parameters such as bond lengths and bond angles are modified efficiently.

Topotactic exchange reactions can be either partial (Fig.~\ref{fig:Synthesis}a) or complete (Fig.~\ref{fig:Synthesis}b).
The most general formulation of a partial exchange reaction is
\begin{equation}
\label{eq:topo1}
 \mathrm {2A_2MO_3+3BX}~\rightarrow~\mathrm{B_3AM_2O_6+3AX}
\end{equation}
where the inter-layer A-atoms (typically Li or Na) in a stable honeycomb structure A$_2$MO$_3$ are exchanged with the B-atoms (typically Cu, Ag, and H) from a halide, nitrate, or sulfate compound BX.
For example, Fig.~\ref{fig:Synthesis}a corresponds to A=Li, B=Ag, M=Ir, and X=NO$_3$ for the synthesis of \ALIO\ from \LIO.
Replacing the inter-layer Li atoms by H, Cu, or Ag, in \LIO\ has recently produced \HLIO, \CLIO, and \ALIO, respectively~\cite{PRL_Bahrami_Thermodynamic, PRB_Geirhos_Quantum, dalton_roudebush_structure}.

In a complete topotactic exchange reaction, all A-atoms within and between the layers are replaced by the B-atoms.
\begin{equation}
\label{eq:topo2}
 \mathrm {A_2MO_3+2BX}~\rightarrow~\mathrm{B_2MO_3+2AX}
\end{equation}
For example, Fig.~\ref{fig:Synthesis}b corresponds to A=Na, B=Cu, M=Ir, and X=Cl for the synthesis of \CIO\ from \NIO.
A complete exchange reaction is much less likely to happen and so far \CIO\ is the only known system in this category~\cite{JACS_abramchuk_cu2iro3}.
It is noteworthy that the copper atoms in \CIO\ are not entirely in a Cu$^+$ state.
Both x-ray absorption and electron energy loss spectroscopy (XAS and EELS) confirmed a mixed valence of Cu$^{+}$/Cu$^{2+}$ with a 1/1 ratio within the honeycomb layers~\cite{kenney_coexistence_2019}. 
A mixed valence of copper, induces a mixed valence of iridium (Ir$^{+3}$/Ir$^{+4}$) and leads to magnetic disorder and spin-glass behavior~\cite{kenney_coexistence_2019,choi_exotic_2019}.

\section{\label{sec:synthesis}Synthesis Details}
\begin{figure}[ht]
\includegraphics[width=0.45\textwidth]{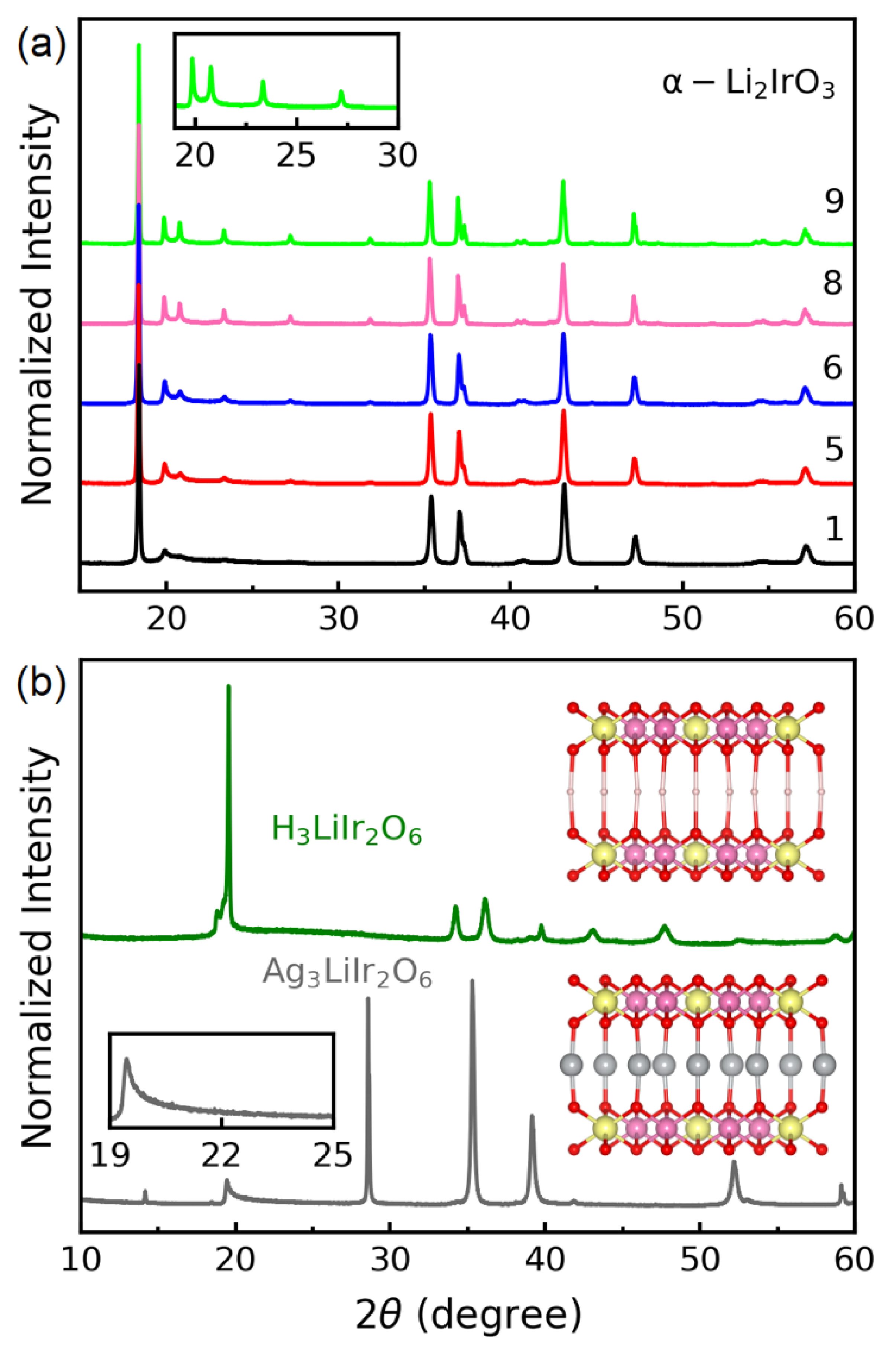}
\caption{\label{fig:XRD}
(a) After each heat cycle, the powder x-ray pattern of \LIO\ shows more pronounced peaks, especially between 20 and 30 degrees where the honeycomb Bragg peaks appear.
The number of times each sample has been reheated is shown on the right above its respective pattern.
(b) The x-ray patterns of two second-generation Kitaev systems, \HLIO\ (green) and \ALIO\ (gray data, reproduced from~\cite{PRB_Bahrami_Effect}). 
The inset shows the asymmetric broadening of the honeycomb Bragg peaks in \ALIO\ due to stacking faults. 
In \HLIO, the honeycomb peaks are hardly discernible due to high structural disorder.
}
\end{figure}
The first-generation Kitaev magnets are prepared via conventional solid state reaction at high temperatures (T$\ge 700$ K) either in air or under the flow of oxygen/argon gas~\cite{PRB_mehlawat_heat_2017, Todorova_LRO_2011, omalley_structure_2008}. 
To improve the sample quality and remove stacking faults, it is necessary to perform successive stages of grinding and heating. 
For example, the x-ray patterns in Fig.~\ref{fig:XRD}a show that the quality of \LIO\ samples improve by repeating the heat cycles. 
Specifically, the superstructure peaks between 20 and 30 degrees (inset of Fig.~\ref{fig:XRD}a) that represent the honeycomb ordering become more pronounced in each iteration.
Typically, improving the quality of the first-generation compound will improve the quality of the second-generation material after the exchange reaction~\cite{PRB_Bahrami_Effect}.

The topotactic cation exchange reaction must be conducted at low temperatures ($T\le 400$~K)~\cite{kitagawa_spinorbital-entangled_2018, JACS_abramchuk_cu2iro3, PRB_Bahrami_Effect}, since higher temperatures will decompose the metastable product. 
The second-generation Kitaev magnets are prepared by modifying the inter-layer atoms and the associated chemical bonds, and therefore they have more stacking faults than their parent compounds~\cite{PRB_Bahrami_Effect, Tsirlin_review_Kitaev_2021}. 
This can be seen in the inset of Fig.~\ref{fig:XRD}b that shows an asymmetric broadening of the honeycomb Bragg peaks in \ALIO.
Unlike the solid state reactions, topotactic exchange cannot be repeated to improve the sample quality.
Thus, getting rid of the stacking faults in these materials remains an open challenge.

Details of the synthesis procedures for \CIO\ and \ALIO\ have previously been published by Abramchuk and Bahrami \emph{et al.}~\cite{JACS_abramchuk_cu2iro3,PRB_Bahrami_Effect,PRL_Bahrami_Thermodynamic}.
Here, we present more details about the synthesis of \HLIO\ based on the earlier work of Bette \emph{et al.}~\cite{bette_solution_2017}.
Polycrystalline samples of \HLIO\ are synthesized using a modified version of Eq.~\ref{eq:topo1}.
\begin{equation}
\label{eq:topo3}
 \mathrm {4Li_2IrO_3+3H_2SO_4}~\rightarrow~\mathrm{2H_3LiIr_2O_6+3Li_2SO_4}
\end{equation} 
After synthesizing a high-quality of \LIO\ (Fig.~\ref{fig:XRD}a), approximately 300~mg of the material was added to a 10~ml Teflon-lined steel autoclave filled with H$_{2}$SO$_{4}$ acid (1 M solution) and heated to 120 $^{\circ}$C for several days. 
After completing the reaction, the product was washed with water and the quality was verified using x-ray diffraction (Fig.~\ref{fig:XRD}b).

\section{\label{sec:FAULT}Stacking Faults}
\begin{figure}[ht]
\includegraphics[width=0.45\textwidth]{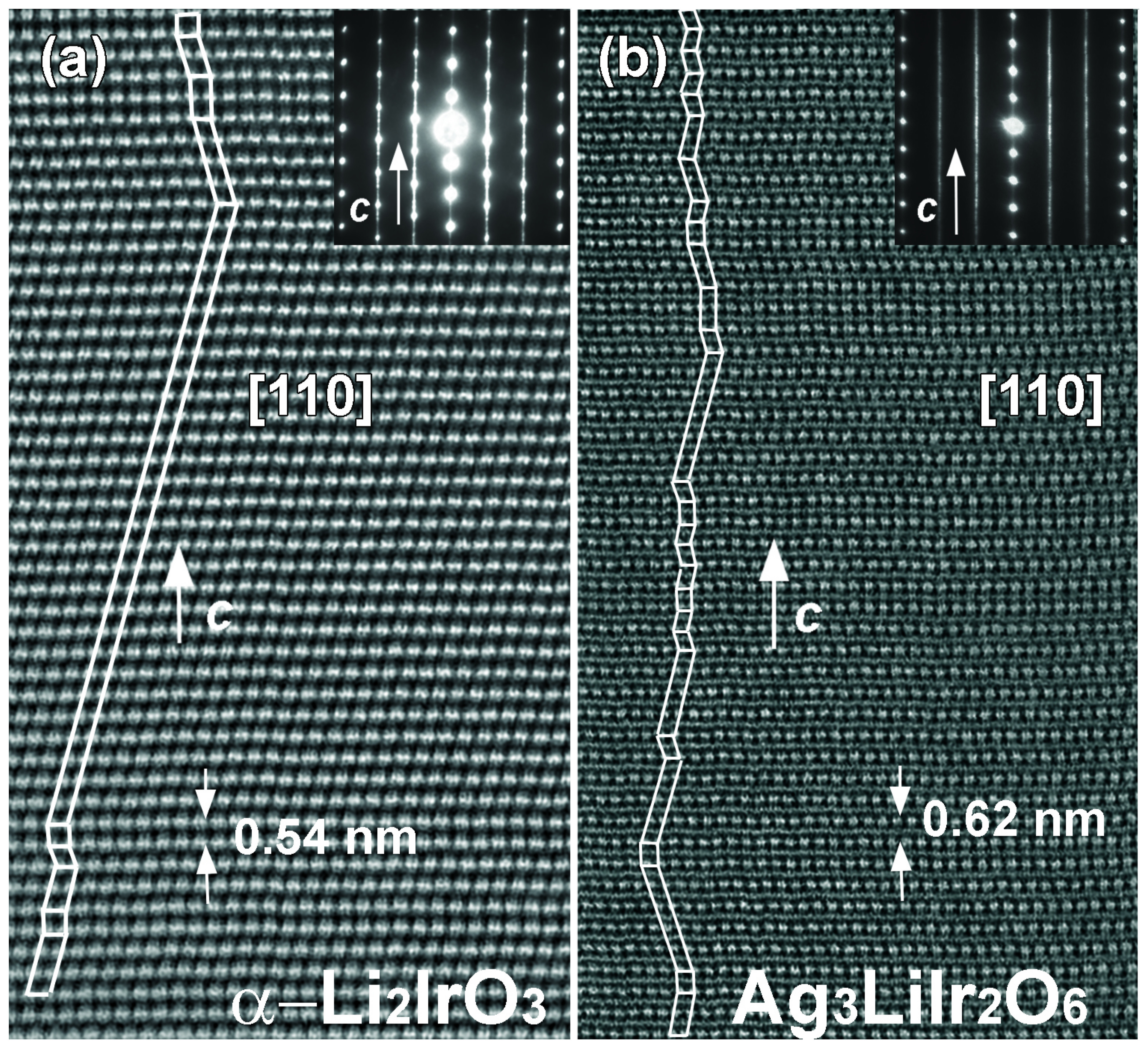}
\caption{\label{fig:TEM}
HAADF-TEM images from (a) \LIO\ and (b) \ALIO.
The images show an abundance of stacking faults in \ALIO\ unlike \LIO, due to the weaker inter-layer bonding in the former.
The electron diffraction patterns are presented as insets and reveal fewer streaking in \LIO\ due to fewer stacking faults compared to \ALIO.
}
\end{figure}
A comparison between the insets of Figs.~\ref{fig:XRD}a and \ref{fig:XRD}b suggests fewer stacking faults in \LIO\ (sharp and well-separated Bragg peaks from the honeycomb layers) and considerable stacking faults in \ALIO\ (broadened peaks).
The asymmetric broadening of honeycomb peaks is known as the Warren line shape, which is a signature of stacking disorder~\cite{balzar_x-ray_1993}.
The higher amount of stacking faults in the second-generation Kitaev magnets is due to the inter-layer chemistry.
As seen in Fig.~\ref{fig:Synthesis}, each inter-layer Li atom in \LIO\ is octahedrally coordinated with three oxygen atoms from the top and three from the bottom honeycomb layers.
In contrast, each Ag atom in \ALIO\ is connected to only one O from the top and one from the bottom layer in a dumbbell (linear) coordination.
The weak dumbbell bonds are responsible for the larger inter-layer separation in \ALIO\ and more stacking faults compared to \LIO~\cite{abramchuk_crystal_2018}.

Direct lattice imaging with transmission electron microscope (TEM) is a powerful tool to study the stacking faults. 
Figures~\ref{fig:TEM}a and \ref{fig:TEM}b (reproduced from Ref.~\cite{PRB_Bahrami_Effect}) are high angle annular dark field TEM (HAADF-TEM) images of \LIO\ and \ALIO\ samples, respectively.
Whereas the stacking sequence in \LIO\ can be flawless for up to 50 unit cells, \ALIO\ shows a maximum of 5 unit cells stacked without faults (in the form of twisting between the layers).
In \HLIO, the small size of H atoms and their high mobility make the chemical bonds even weaker than in \ALIO. 
As such, \HLIO\ has the highest degree of stacking faults among the second-generation Kitaev magnets~\cite{bette_solution_2017,kitagawa_spinorbital-entangled_2018,Tsirlin_review_Kitaev_2021}.
This is why the honeycomb peaks of \HLIO\ are not resolved by x-rays (Fig.~\ref{fig:XRD}b).

\section{\label{sec:TUNING}Tuning magnetic interactions with topochemical methods}
\begin{figure}[ht]
\includegraphics[width=0.45\textwidth]{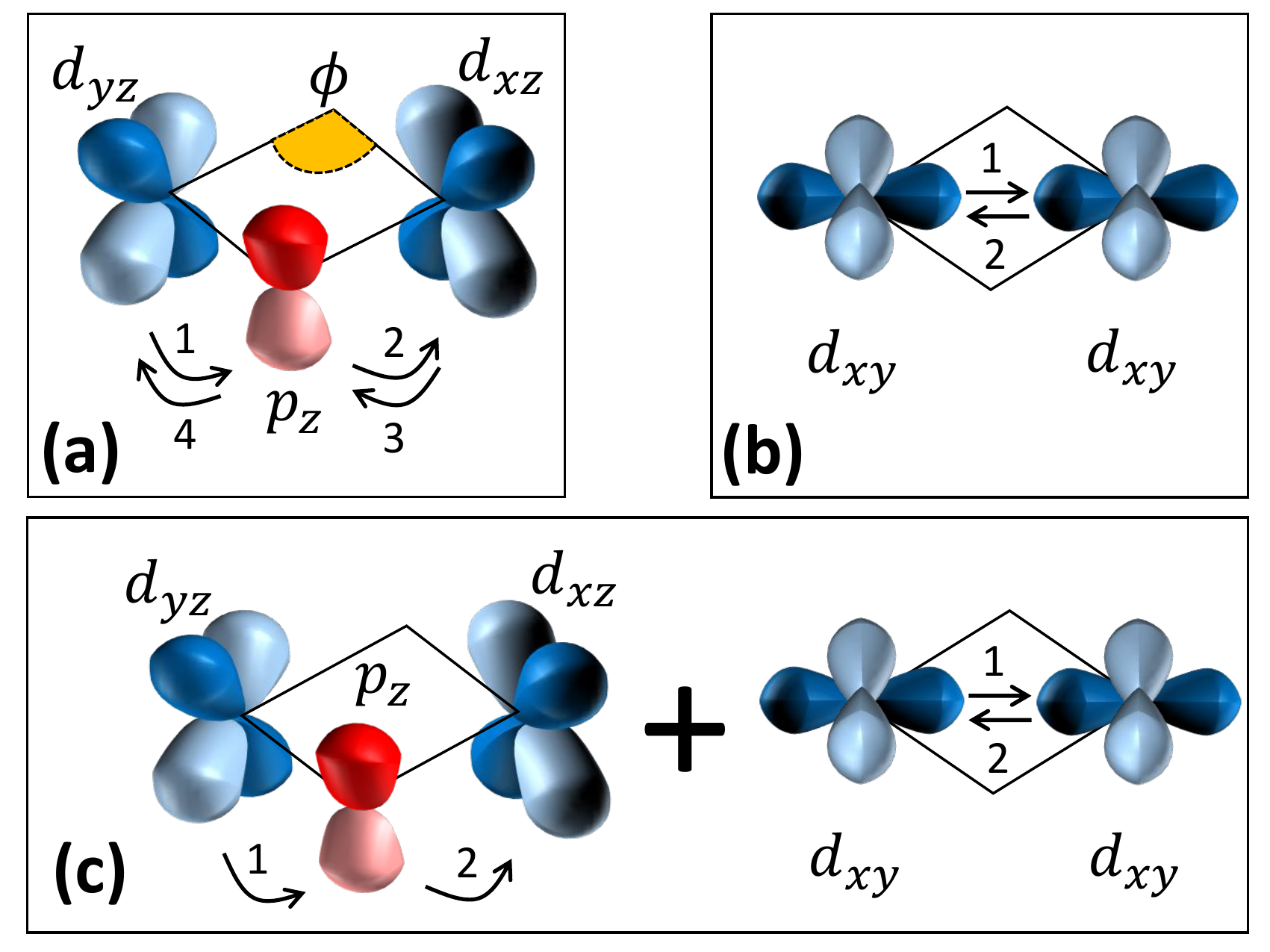}
\caption{\label{fig:ORBS}
Exchange paths for (a) $K$, (b) $J$, and (c) $\Gamma$ terms in Eq.~\ref{eq:Kit}.
The $d$ and $p$ orbitals are painted in blue and red, respectively.
The numbers show the hopping sequence in the perturbation.
}
\end{figure}
As seen in Fig.~\ref{fig:Synthesis}, the monoclinic unit cell and the honeycomb ordering in the 2D layers remain unchanged before and after exchange reactions.
However, the change of inter-layer coordination from octahedral to dumbbell modifies the M-O-M bond angles within the honeycomb layers (Fig.~\ref{fig:T2G}a and \ref{fig:Synthesis}).
Superexchange magnetic interactions are sensitive to a change of bond angles and thus topochemical reactions can be used to tune the magnetic interactions. 
There are at least three terms in the magnetic Hamiltonian of the Kitaev materials.
\begin{equation}
\label{eq:Kit}
\mathcal{H}=\sum\limits_{\langle i,j \rangle \in {\alpha\beta(\gamma)}} \left[ -K_{\gamma}S_i^{\gamma}S_j^{\gamma} + J \textbf{S}_i  \cdot \textbf{S}_j+ \Gamma \left( S_i^{\alpha}S_j^{\beta} + S_i^{\beta}S_j^{\alpha}\right) \right]
\end{equation}
The Kitaev term ($K$) favors QSL, the Heisenberg term ($J$) favors AFM ordering, and the off-diagonal exchange term ($\Gamma$) controls details of the ordered structure.
All three term can be modified via topochemical reactions as follows.

Figure~\ref{fig:ORBS} shows the individual exchange paths for each term in Eq.~\ref{eq:Kit}.
The Kitaev term is an indirect exchange interaction with hopping matrix elements $t_{dpd}$ between the $d_{xz}$, $p_z$, and $d_{yz}$ orbitals (Fig.~\ref{fig:ORBS}a)~\cite{rau_spin-orbit_2016,kim_crystal_2016}.
In addition to the indirect exchange ($K$), Fig.~\ref{fig:ORBS}b shows a direct exchange path for the Heisenberg interaction ($J$) with hopping matrix element $t_{dd}$ between $d_{xy}$ orbitals, leading to $J\sim t_{dd}^2/U$ in Eq.~\ref{eq:Kit}~\cite{winter_challenges_2016}.
Finally, a combination of direct and indirect paths in Fig.~\ref{fig:ORBS}c leads to the symmetric off-diagonal exchange, $\Gamma\sim t_{dpd}t_{dd}J_H/U^2$, where $J_H$ is the Hund's coupling between the \eg\ and \tg\ orbitals~\cite{PRL_Rau_Generic,rusnacko_kitaev-like_2019}.
The hopping matrix elements ($t_{dd}$ and $t_{dpd}$) are tuned by the M-O-M bond angle and the M-M distance which can be tuned by the exchange reactions.
For example, (i) the change of oxygen positions within the honeycomb layers due to the change of inter-layer coordinations in Fig.~\ref{fig:Synthesis} modifies the M-O-M bond angle ($\phi$ in Fig.~\ref{fig:T2G}a and \ref{fig:ORBS}a) and thereby tunes $t_{dpd}$; 
(ii) according to theoreical calculations~\cite{PRL_Jackeli_Mott}, the Heisenberg interaction is canceled between the opposite paths if the bond angle $\phi$ is close to $90^\circ$ (Fig.~\ref{fig:T2G}a and \ref{fig:ORBS}a);
(iii) the hybridization between the Ag $d$-orbitals between the layers and O $p$-orbitals within the layers tunes the ratio of $t_{dpd}/t_{dd}$.

\section{\label{sec:MAGNETISM}Magnetic characterization of metastable Kitaev materials}
\begin{figure}[ht]
\includegraphics[width=0.45\textwidth]{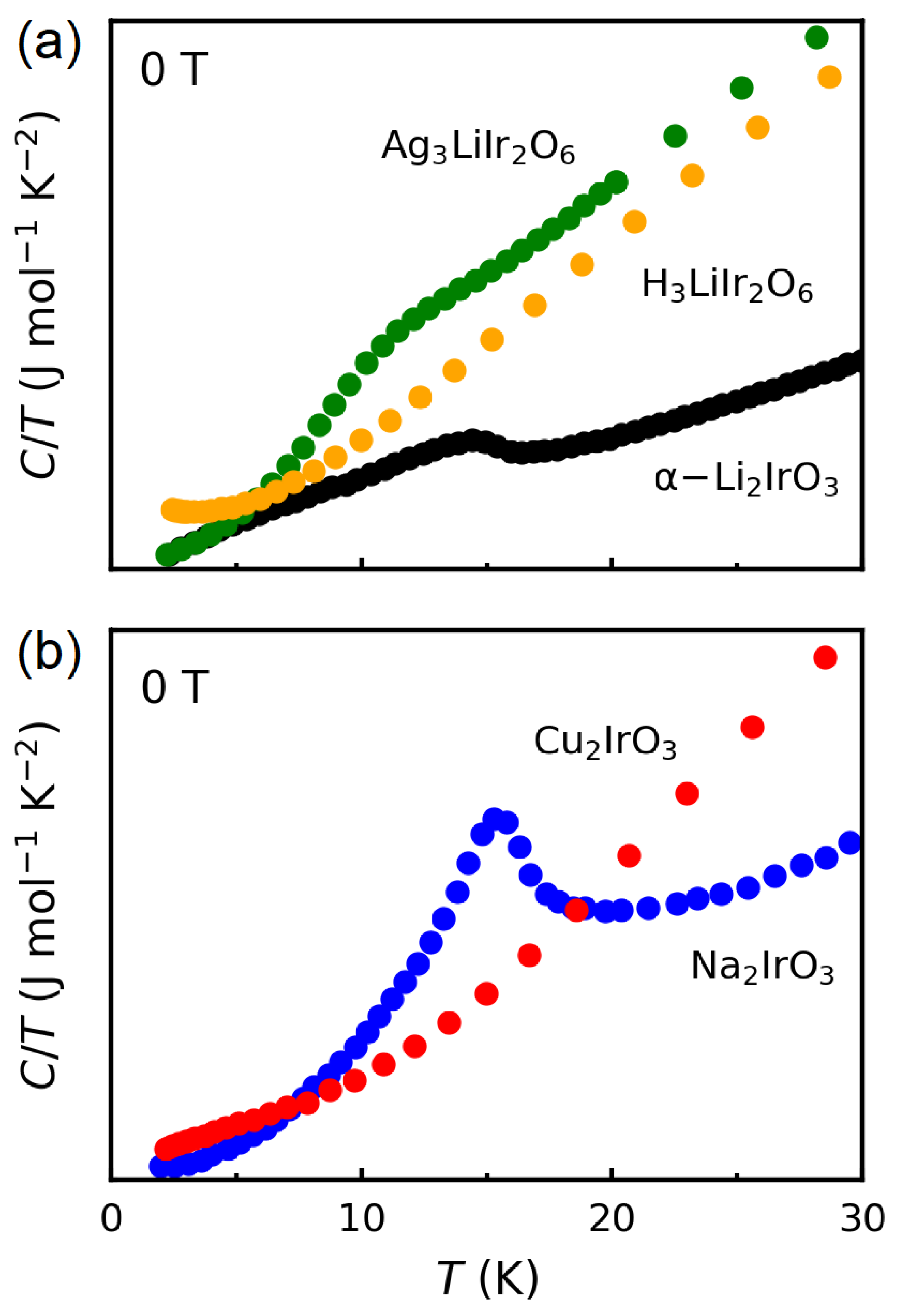}
\caption{\label{fig:CP}
(a) Heat capacity ($C/T$) plotted as a function of temperature below 30~K for the first-generation Kitaev magnet \LIO\ and its second-generation derivatives \ALIO\ and \HLIO.
The data for \LIO\ and \ALIO\ are reproduced from Refs.~\cite{PRL_Singh_A2IrO3,PRB_Bahrami_Effect}.
(b) A similar comparison is made between \NIO\ (first-generation) and \CIO\ (second-generation).
The data are reproduced from Ref.~\cite{JACS_abramchuk_cu2iro3}.
}
\end{figure}
To demonstrate the effect of topochemical modifications on the magnetic interactions (Eq.~\ref{eq:Kit} and Fig.~\ref{fig:ORBS}), we compare the heat capacity and magnetic susceptibility of the first and second-generation Kitaev magnets.
The peak in the heat capacity of \LIO\ in Fig.~\ref{fig:CP}a confirms long-range magnetic ordering at $T_N=15$~K.
The order has been characterized as incommensurate spiral by recent neutron scattering and muon spin relaxation ($\mu$SR) experiments~\cite{PRB_Williams_Incommensurate,PRB_Choi_Spin}.
As seen in Fig.~\ref{fig:CP}a, this peak is shifted to lower temperatures in \ALIO\ and seemingly disappeared in \HLIO.
The suppression of $T_N$ in second-generation compounds \ALIO\ and \HLIO\ is a positive sign of approaching the quantum spin-liquid (QSL) phase, where long-range order is replaced by long-range quantum entanglement~\cite{Nature_takagi_concept_2019,knolle_field_2019}.
A recent $\mu$SR experiment~\cite{PRB_Bahrami_Effect} has shown a similar incommensurate spiral order in \ALIO; however, the long-range order develops at 8~K in \ALIO, well below $T_N=15$~K in \LIO.
Thus, the topochemical modification of bond angles seem to strengthen $K$ and weaken $J$ in Eq.~\ref{eq:Kit}.
A recent nuclear magnetic resonance (NMR) experiment has shown absence of long-range order in \HLIO, which is another promising result toward the discovery of a QSL phase~\cite{kitagawa_spinorbital-entangled_2018}.

A similar trend is observed in Fig.~\ref{fig:CP}b for the first-generation material \NIO\ that shows a peak at $T_N=15$~K and its second-generation counterpart \CIO\ that does not show a peak but seems to have a broad anomaly below 5~K.
Neutron scattering experiments have confirmed a zigzag antiferromagnetic (AFM) order in \NIO~\cite{choi_spin_2012}.
Recent $\mu$SR and NMR experiments have revealed a coexistence of static and dynamic magnetism below 5~K in \CIO\ but without a long-range order, suggesting proximity to the QSL phase~\cite{kenney_coexistence_2019,takahashi_spin_2019}.

\begin{figure}[ht]
\includegraphics[width=0.45\textwidth]{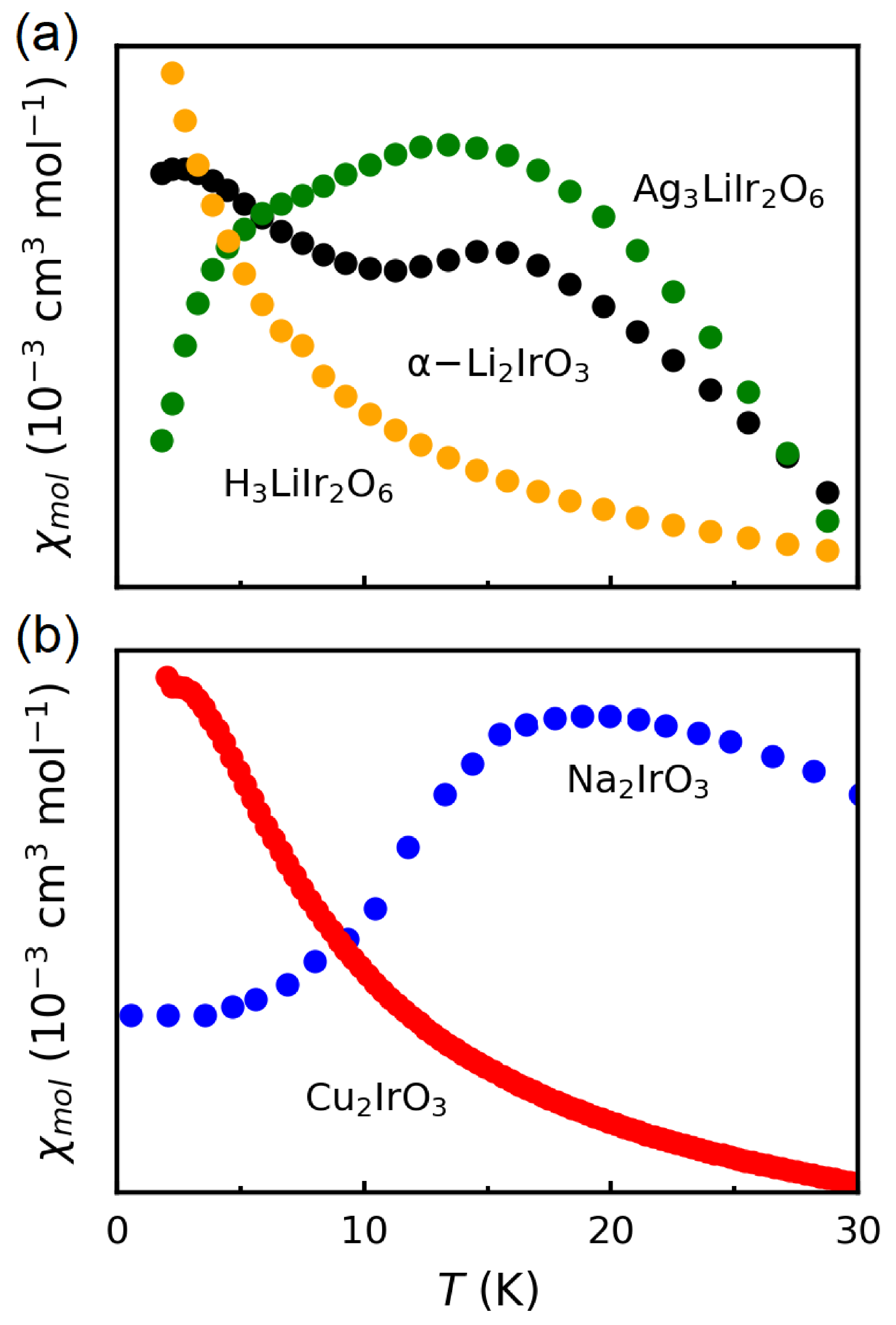}
\caption{\label{fig:XT}
(a) Magnetic susceptibility ($\chi$) plotted as a function of temperature below 30~K for the first-generation Kitaev magnet \LIO\ and its second-generation derivatives \ALIO\ and \HLIO.
The data for \LIO\ and \ALIO\ are reproduced from Refs.~\cite{PRL_Bahrami_Thermodynamic,PRB_Bahrami_Effect}.
(b) A similar comparison is made between \NIO\ (first-generation) and \CIO\ (second-generation).
The data for \NIO\ and \CIO\ are reproduced from Refs.~\cite{PRL_Singh_A2IrO3,JACS_abramchuk_cu2iro3}.
}
\end{figure}
The suppression of magnetic ordering due to topochemcial changes in metastable Kitaev magnets is also observed in the magnetic susceptibility data.
Figure~\ref{fig:XT}a shows the magnetic susceptibility of \LIO\ (black curve) with a clear anomaly at $T_{N} = 15$~K indicating the incommensurate spiral antiferromagnetic (AFM) order. 
The green curve representing \ALIO\ shows two downturns at $T_{F}=14$~K and $T_{N}=8$~K, corresponding to the onsets of short-range and long-range magnetic orders, respectively~\cite{PRB_Bahrami_Effect}. 
The orange curve representing \HLIO\ does not show any evidence of magnetic ordering.
Figure~\ref{fig:XT}b shows a similar trend, where the first-generation material \NIO\ orders at $T_N=15$~K and the second-generation material \CIO\ shows a small peak at 2~K, evidence of short-range spin freezing instead of long-range order.

\section{\label{sec:Disorder}Challenges and Opportunities}
The above results are exciting; however, they need to be interpreted with caution.
Topotactic exchange reactions increase disorder that has adverse effects on magnetism.
A recent TEM study has shown that the silver atoms in \ALIO\ can enter the honeycomb layers and form small inclusions (up to 50 atoms) that disrupt the magnetic ordering~\cite{PRB_Bahrami_Effect}.
Such a structural disorder can spuriously hide the long-range order and be misinterpreted as evidence of a QSL phase. 
As noted earlier, \HLIO\ is even more disordered compared to \ALIO\ due to the high mobility of the H atoms, which causes bond randomness and site vacancies within the honeycomb layers~\cite{bette_solution_2017}.    
Recent theoretical works show that the absence of magnetic ordering in \HLIO\ may be due to bond randomness and a large amount of vacancies~\cite{knolle_bond-disordered_2019,kao_vacancy-induced_2021}.
Thus, the most important challenge in this field is to optimize the synthesis conditions for a minimum amount of disorder and to find methods of annealing away the stacking faults and vacancies.

Metastable Kitaev magnets have opened a new window of opportunity to realizing the quantum spin-liquid ground state.
The Majorana excitations of such materials will form the building blocks of a solid state quantum computer~\cite{frolov_quantum_2021}.
Braiding algorithms and logical gates have been theoretically developed for such computers~\cite{field_introduction_2018}.
It remains an open challenge for the solid state chemistry community to synthesize the appropriate materials for such models.
Another intriguing opportunity is to find unconventional superconductivity in the Kitaev magnets~\cite{you_doping_2012}, an exciting theoretical prediction that awaits experimental discovery.





\section{\label{sec:ackn}ACKNOWLEDGMENTS}
This material is based upon work supported by the Air Force Office of Scientific Research under award number FA2386-21-1-4059.

\bibliography{Bahrami_20Dec2021}
\end{document}